%% file: main.tex
\documentclass[fleqn,usenatbib]{mnras}

\usepackage{eso-pic}

\AddToShipoutPictureBG*{%
  \AtPageUpperLeft{%
    \hspace{0.75\paperwidth}%
    \raisebox{-1.\baselineskip}{%
      \makebox[0pt][l]{\textnormal{DES-2015-0048}}
}}}%

\AddToShipoutPictureBG*{%
  \AtPageUpperLeft{%
    \hspace{0.75\paperwidth}%
    \raisebox{-2.\baselineskip}{%
      \makebox[0pt][l]{\textnormal{FERMILAB-PUB-23-369}}
}}}%

\usepackage{newtxtext,newtxmath}
\usepackage{mdframed}
\usepackage{graphicx}
\usepackage{subcaption}
\usepackage{lineno}
\usepackage{dsfont}
\let\oldequation\equation
\let\oldendequation\endequation
\renewenvironment{equation}
  {\linenomathNonumbers\oldequation}
  {\oldendequation\endlinenomath}

\usepackage{graphicx}
\usepackage{float}
\usepackage{hyperref}
\usepackage[T1]{fontenc}
\usepackage{ae,aecompl}
\usepackage{todonotes}

\usepackage{float}
\usepackage{graphicx}	
\usepackage{amsmath}	
\usepackage{adjustbox}
\usepackage{multirow}

\newcommand{\avg}[1]{\langle #1 \rangle}

\newcommand{\minorsection}[1]{\vspace{0.1cm} \noindent \textbf{{#1}}:}

\definecolor{purple}{RGB}{150,0,200}

\usepackage{lineno}

\title[Source clustering in DES Y3]{Detection of the significant impact of source clustering on higher-order statistics with DES Year 3 weak gravitational lensing data}

\makeatletter
\def \blfootnote{\xdef\@thefnmark{}\@footnotetext}
\makeatother

\input{authors_MNRAS.tex}		
\date{\today}

\begin{document}     
\label{firstpage} 
\pagerange{\pageref{firstpage}--\pageref{lastpage}}%
\maketitle        	 

\begin{abstract}
We demonstrate and measure the impact of source galaxy clustering on higher-order summary statistics of weak gravitational lensing data. By comparing simulated data with galaxies that either trace or do not trace the underlying density field, we show this effect can exceed measurement uncertainties for common higher-order statistics for certain analysis choices. Source clustering effects are larger at small scales and for statistics applied to combinations of low and high redshift samples, and diminish at high redshift. We evaluate the impact on different weak lensing observables, finding that third moments and wavelet phase harmonics are more affected than peak count statistics. Using Dark Energy Survey Year 3 data we construct null tests for the source-clustering-free case, finding a $p$-value of $p=4\times10^{-3}$ (2.6 $\sigma$) using third-order map moments and $p=3\times10^{-11}$ (6.5 $\sigma$) using wavelet phase harmonics. The impact of source clustering on cosmological inference can be either be included in the model or minimized through \textit{ad-hoc} procedures (e.g. scale cuts). We verify that the procedures adopted in existing DES Y3 cosmological analyses (using map moments and peaks) were sufficient to render this effect negligible.  Failing to account for source clustering can significantly impact cosmological inference from higher-order gravitational lensing statistics, e.g. higher-order N-point functions, wavelet-moment observables (including phase harmonics and scattering transforms), and deep learning or field level summary statistics of weak lensing maps. We provide recipes both to minimise the impact of source clustering and to incorporate source clustering effects into forward-modelled mock data. 
\end{abstract}






\section{Introduction}

Weak gravitational lensing from large-scale structure in the Universe induces small distortions in the observed shape of background source galaxies. The weak lensing signal can be measured using large samples of galaxies to observe correlated distortions in observed galaxy ellipticities \citep[see][]{Bartelmann2001}. 
The angular distribution of source galaxies is not uniform; it is modulated by observational and selection effects (such as varying observing depth) and by clustering due to galaxies tracing the underlying density field. The latter effect, called \textit{source clustering} \citep{Schneider2002,Schmidt2009,Valageas2014,Krause2021}, causes the galaxy number density to be correlated with the target lensing signal: since we expect a larger lensing signal along overdense lines-of-sight, we preferentially sample the shear field where its value is larger. 
For pixelized shear maps, this results in two distinct effects: (1) the average noise-free lensing signal is modulated by a different effective redshift distribution, and (2) the \textit{shape noise} (due to the intrinsic ellipticities of galaxies) is correlated with the lensing signal.

Higher-order statistics have recently been growing in popularity as powerful tools for efficiently extracting cosmological information from current weak lensing data (e.g. \citealt{ Liu2015,Vicinanza2016,Martinet2018, Fluri2019,Cheng2020,moments2021,jeffrey_lfi,Zuercher2022,Lu2023}).  Their use can improve constraints on cosmological parameters (relative to standard two-point statistics), can help discriminate between general relativity and modified gravity theories \citep{Cardone2013,Peel2018}, and can help self-calibrate astrophysical and observational nuisance parameters \citep{Pyne2021}. Given the increasing precision of these measurements, the impact of systematic errors on higher-order statistics is a subject of careful consideration.
 
 The impact of source clustering has generally been neglected in the forward model, as it has often been considered a small, higher-order contribution to weak lensing observables. The efficiency of lensing peaks roughly halfway between the source and the observer, and vanishes at the source location; any correlation between the shear field `seen' by the source galaxy and the density field it lives on is suppressed. Source clustering has been studied in the context of two-point correlation functions, and theoretical calculations by \cite{Krause2021} have shown it to be negligible for Stage III surveys for catalogue-based Gaussian statistics. Whether its impact on weak lensing higher-order statistics is also negligible is less clear, although some early estimates suggested a stronger impact on three-point correlation functions \citep{Valageas2014}. The effect of source clustering has not to date been explicitly included in the suites of simulations used for simulation-based cosmological analyses \citep[e.g.][]{Martinet2018,Zuercher2021}, although peak statistics analyses by \cite{Kacprzak2016} and by \cite{Zuercher2021} performed initial tests of this effect (under some simplifying assumptions), showing no significant effect on their cosmological constraints.

This work develops a forward-modelling procedure to introduce source clustering effects into the simulated maps.  
We consider the impact of source clustering on several non-Gaussian observables, looking primarily at map-based estimators.
We show that source clustering generates a clear signature on higher-order summary statistics for specific analysis choices, we demonstrate this effect in the Dark Energy Survey (DES) Year 3 (Y3) data, and we discuss the impact of this effect on previously published DES measurements.
\section{Data and Simulations}
\subsection{DES Y3 weak lensing catalogue}
We use the DES Y3 weak lensing catalogue \citep*{y3-shapecatalog}; this consists of 100,204,026 galaxies, with a weighted $n_{\rm eff}=5.59$~galaxies~arcmin$^{-2}$, over an effective area of 4139  deg$^2$. It was created using the \textsc{METACALIBRATION} algorithm \citep{HuffMcal2017, SheldonMcal2017}, which provides self-calibrated shear estimates starting from (multi-band) noisy images of the detected objects. A residual small calibration (via a multiplicative shear bias) is provided; based on sophisticated image simulations \citep{y3-imagesims}, it accounts for blending-related detection effects. An inverse variance weight is further assigned to each galaxy in the catalogue to enhance the overall signal-to-noise. The sample is divided into four tomographic bins of roughly equal number density \citep*{y3-sompz}. Redshift distributions are provided by the SOMPZ method, in combination with  clustering redshift constraints \citep*{y3-sompz}. 

\vspace{-0.1cm}
\subsection{Simulations}\vspace{-0.1cm}
We rely on simulations produced using the \textsc{PKDGRAV3} code \citep{potter2017pkdgrav3}. We use 50 independent realisations at the fixed cosmology $\Omega_{\rm m} = 0.26$, $\sigma_8 = 0.84$, $\Omega_{\rm b} = 0.0493$, $n_{\rm s} = 0.9649 $, $h = 0.673$ from the DarkGridV1 suite, described in detail in \cite{Zuercher2021,Zuercher2021b}. All simulations include three massive neutrino species with a mass of $m_{\nu}=0.02$ eV per species. The simulations were obtained using 14 replicated boxes in each direction ($14^3$ replicas in total) so as to span the redshift interval from $z = 0$ to $z = 3$. Each individual box contains $768^3$ particles and has a side-length of 900 $h^{ - 1}$ Mpc.  For each simulation, lens planes $\delta_{\rm shell}(\hat{\boldsymbol{\rm n}}, \chi)$ are provided at $\sim87$ redshifts from $z=3$ to $z=0$. The lens planes are provided as \textsc{HEALPIX} \citep{GORSKI2005} maps and are obtained as the overdensity of raw number particle counts; for this work, we downsample the orginal resolution of \textsc{NSIDE} = 2048 to \textsc{NSIDE} = 1024 (with pixel size $\approx$ 3.4 arcmin).  The lens planes are converted into convergence planes $\kappa_{\rm shell}(\hat{\boldsymbol{\rm n}}, \chi)$ under the Born approximation (e.g. Eq. 2 from \citealt{Fosalba2015}). Lastly, shear planes $\gamma_{\rm shell}(\hat{\boldsymbol{\rm n}}, \chi)$ are obtained from the convergence maps using a full-sky generalisation of the \cite{KaiserSquires} algorithm \citep*{y3-massmapping}.


\section{Source Clustering Implementation}\label{sec:implementation}
In the limit of high source galaxy density, the observed projected shear in direction $\vv{\theta}$ will be
\begin{equation}
\gamma(\vv{\theta}) = \frac{\int n(\vv{\theta}, z) \, \gamma(\vv{\theta}, z) \, \mathrm{d} z}{\int n(\vv{\theta}, z) \, \mathrm{d} z},
\end{equation}
\noindent where $n(\vv{\theta}, z)$ is the unnormalised galaxy density (i.e. $\int_{V} n(\vv{\theta}, z) \, \mathrm{d} \! \vv{\theta} \mathrm{d} z$ is the number of source galaxies in the volume $V$). The observed shear $\gamma$ is the sum of signal $\gamma_s$ and noise $\epsilon_n$:
\begin{multline}
\gamma(\vv{\theta}) =
\frac{\int n(\vv{\theta}, z) \Big( \gamma_s(\vv{\theta}, z) + \epsilon_n(\vv{\theta}, z) \Big) \, \mathrm{d}z}{\int n(\vv{\theta}, z) \, \mathrm{d} z} = \gamma_{s}(\vv{\theta}) + \gamma_{n}(\vv{\theta}) .
\end{multline}
It has been standard in many previous analyses to use the spatial average
\begin{equation}
\bar{n}(z) = \frac{\int n(\vv{\theta}, z) \, \mathrm{d} \! \vv{\theta}}{\int \mathrm{d} \! \vv{\theta}}
\end{equation}
as an approximation to $n(\vv{\theta}, z)$; however, this approximation cannot include the effect of source clustering. 
We instead model the directional variation of the source galaxy distribution arising from its dependence on the overdensity field $\delta(\vv{\theta}, z)$, i.e. $n(\vv{\theta}, z) = \bar{n}(z) \left[ 1 + f(\delta(\vv{\theta}, z)) \right]$ for some function $f$. This leads to a relation between $n(\vv{\theta}, z)$ and the observed shear $\gamma(\vv{\theta}, z)$, as they both depend on $\delta$.
This relation has a direct impact on the expected value $\gamma_{s}$ (i.e. the signal is modulated). 
Additionally, as the variance of the noise term $\gamma_{n}$ depends on $n$ (more source galaxies leads to reduced noise), this relation will have an impact on the expected value of terms such as $\gamma_{s}\gamma_{n}^2$.
A simulation that does not include source clustering effects is in danger of incorrectly modelling these expected values.


Below we describe how to create pixelized shear maps both without and with source clustering effects. We consider one fixed tomographic bin. We assume as inputs a noiseless pixelized simulated shear map and a separate galaxy shape catalogue. The latter is needed to supply shape noise information (as the simulated shear map is not assumed to have an associated simulated galaxy catalogue); in our case the DES Y3 shape catalogue serves this purpose. We then add a source clustering effect by amending both signal and noise terms using factors related to the matter overdensity in the shear simulation.

An alternative method for creating shear simulations with source clustering would be to use the results of the n-body simulation (i.e. the simulation used to create the simulated shear field) to directly create a galaxy catalogue (using some HOD prescription, for example), to assign shape noise to these galaxies, and to use this information to add noise to the shear simulation. However this task is complex, and therefore we opt for the simpler approach implemented in this work.

 In what follows let $p$ be a pixel, $s$ a thin redshift shell, $\gamma(p, s)$ the noiseless shear from the shear simulation, and $\bar{n}(s)$ the galaxy count across the whole footprint \citep{y3-sompz}. From the galaxy catalogue, let $g$ denote a galaxy, $w_g$ its weight, and $e_g$ its ellipticity after the application of a random rotation to erase the shear signal.

\minorsection{Mock shear maps with no source clustering}
The output simulated shear for a given pixel $p$ is the sum of signal and shape noise contributions:
\begin{equation}
\gamma(p) = \frac{\sum_s \bar{n}(s) \gamma(p, s)}{\sum_s \bar{n}(s)} + \frac{\sum_g w_g e_g}{\sum_g w_g}.
\end{equation}
In the signal term the sum is over all shells $s$, and in the noise term the sum is over all the shape catalogue (i.e. DES Y3) galaxies $g$ in $p$.

\minorsection{Mock shear maps with source clustering}
Let $\delta(p, s)$ be the matter overdensity in the shear simulation. Let $b_g$ be the galaxy-matter bias; for simplicity we assume linear biasing to hold and moreover for our main tests we assume $b_g=1$ (a reasonable choice for the blue field galaxies that constitute most of the galaxies in the shear catalogue). The factor $\bar{n}(s) \left[1 + b_g \delta(p, s)\right]$ is then the relative galaxy count in pixel $p$ and shell $s$; it is generated from the shear simulation and is therefore consistent with the shear signal. In the output simulated shear both the signal and the shape noise contributions have been amended to account for source clustering as follows:
\begin{multline}
\label{eq:sc_pixel}
\gamma_{\textrm{SC}}(p) = \frac{\sum_s \bar{n}(s) \left[1 + b_g \delta(p, s) \right] \gamma(p, s)} {\sum_s \bar{n}(s) \left[1 + b_g \delta(p, s)\right]} \  + \\
\left(\frac{\sum_s \bar{n}(s)}{\sum_s \bar{n}(s) \left[1 + b_g \delta(p, s)\right]}\right)^{1/2} F(p) \, \frac{\sum_g w_g e_g}{\sum_g w_g}.
\end{multline}
The signal term is a weighted average over shells; here the average has been amended to include a shear-correlated source galaxy count. In the shape noise term there are two additional factors. The first, a source clustering factor, results in the shape noise variance scaling as the inverse of the relative galaxy count, as desired; this gives a correlation between the shear signal in a pixel and the square of the shape noise that was not present before. The second, $F(p)$, is a near-unity scale factor introduced to avoid double-counting source clustering effects. The DES Y3 catalogue used to model the shape noise of the pixels is already affected by source clustering. In practice this means that the noise of the catalogue is already modulated by $1/\sqrt{\sum_s \bar{n}(s) \left[1 + b_g \delta_{\rm data}(p, s)\right]}$. This modulation is not correlated with the large scale structure of the simulations. However, since Eq. \ref{eq:sc_pixel} introduces a similar modulation, the net effect is that the even moments of the pixel's simulated noise (variance,  kurtosis, etc.) are slightly enhanced with respect to data, mostly at small scales and low redshifts. The function $F(p)$ corrects this enhancement. We opted for the following expression:
\begin{equation}
    F(p) = A\sqrt{1-B \sigma_{e}^2(p)},
\end{equation}
where the coefficients $A$ and $B$ are per-bin constants, and  
$\sigma_{e}^2(p)$ is the variance of the pixel noise. This correction is (only mildly) cosmology dependent; we used our simulations at fixed cosmology to estimate the two sets of constants for the four bins: $A = [0.97,0.985,0.990,0.995]$, and $B = [0.1,0.05,0.035,0.035]$.


\minorsection{Remarks concerning our implementation}
We generate shear maps for each tomographic bin. The 50 independent simulations at fixed cosmology for our main tests yield 200 independent simulated DES Y3 shear catalogues (as we can cut four independent DES Y3 footprints from each full-sky map). The simulations have not been run at the best-fitting cosmology for the data. However, based on the results presented in \cite{moments2021}, the cosmology chosen for the simulations should still provide a reasonable fit to the data. Moreover, for simplicity we did not include any intrinsic alignments and we assumed zero shear and redshift biases; we do not expect this to affect any of our conclusions.

\section{Results}

\begin{figure}
\begin{center}
\includegraphics[width=0.49 \textwidth]{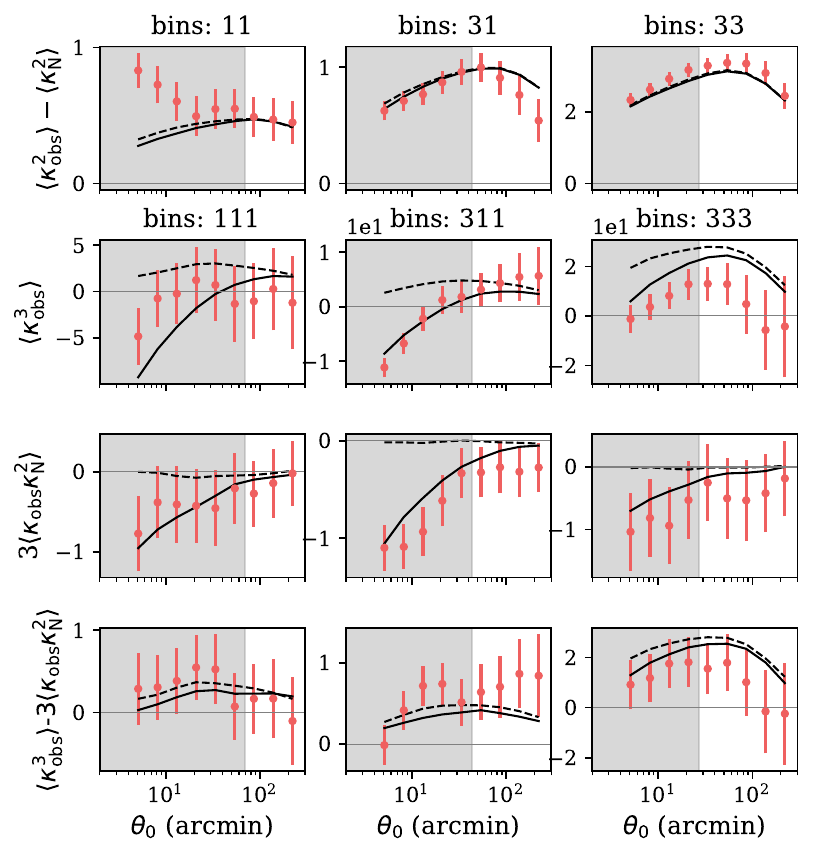}
\includegraphics[width=0.49 \textwidth]{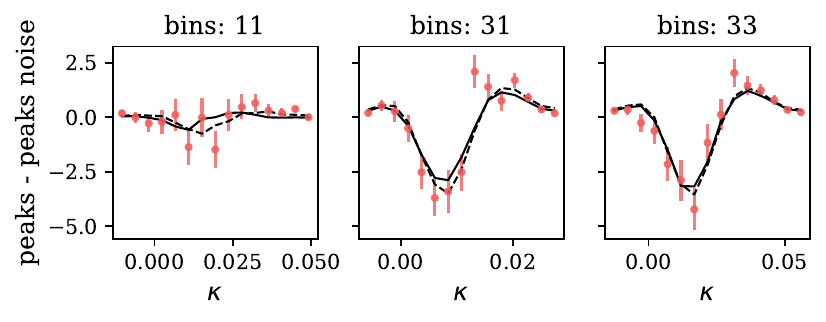}
\includegraphics[width=0.49 \textwidth]{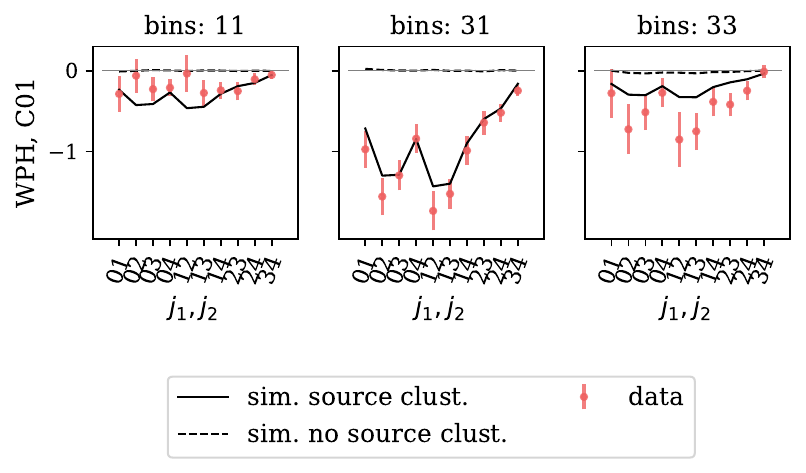}
\end{center}
\caption{From top to bottom: second and third moments, peaks functions, and wavelet phase harmonics for three different combinations of maps involving different tomographic bins (e.g. `bins: 311' means two maps for the first bin and one for the third bin have been used). In each plot, red points represent the measurement in DES Y3 data and solid (resp. dashed) lines represent the average measurements over the set of simulations with (resp. without) source clustering. We multiplied the amplitude of each statistic by a constant to re-scale the dynamical range on the y-axis for plotting purposes. Where present, the grey shaded regions represent the scales that have been not considered in the cosmological analyses using moments \citep{moments2021}. Third moments involving noise are labelled $3 \avg{{\kappa}{\kappa}^2_{{\rm N}}} \equiv \langle \kappa \kappa_{\rm N} \kappa_{\rm N} \rangle^{i,j,k} + {\rm cycl.} $, with `cycl.' referring to the cyclic permutation of the indexes of the bins.}
\label{fig:sc}
\end{figure}


In this work, we consider the following summary statistics:

\begin{itemize}
    \item \textbf{Second and Third Map Moments}: second moments are a Gaussian statistic (i.e. a function only of the power spectrum), whereas third moments probe additional non-Gaussian features of the field \citep{VanWaerbeke2013,Petri2015,Chang2018,Vicinanza2018,Peel2018, G20,moments2021}. Second and third moments of the DES Y3 weak lensing mass maps were used in \cite{moments2021} to infer cosmology; we use the same implementation of the moments estimator.
    \item \textbf{Peaks}: the peaks statistic counts the number of peaks of the smoothed map above a certain threshold. We follow the implementation of peak counts in \cite{Zuercher2021b}. 
    \item \textbf{Wavelet Phase Harmonics (WPH)}: these statistics are part of a broader set of methods (which include \textit{wavelet scattering transforms}, e.g. ~\citealt{Cheng2020}) that were designed to emulate information capture in the manner of a convolutional neural network~\citep{Mallat_2016} without the need for training data. WPH statistics characterise the coherent structures in non-Gaussian random fields, by quantifying the phase alignment at different spatial scales~\citep{Mallat2020, Zhang2019}, and they can provide useful insights as a direct analogy with deep learning. 
    We follow the implementation of WPH in \cite{Allys2020}, which has already found success with astrophysical applications~\citep{bruno_denoising, jeffrey_lfi_wph}.
\end{itemize}

These map-based statistics are applied to reconstructed weak lensing mass maps, using a full-sky generalisation of the \cite{KaiserSquires} algorithm that recovers a noisy estimate of the lensing convergence field $\kappa$ from pixelized shear maps \citep*[see][]{y3-massmapping}. The statistics are applied to `smoothed' versions of the maps. More details about the specific implementation of each statistic is provided in Appendix \ref{sect:summary_stats}. 

{For each statistic, we assess in Fig. \ref{fig:sc} the impact of source clustering by comparing the measurements from the simulations with and without source clustering (solid and dashed lines); these measurements are then compared to data (red points). When possible, we highlight the part of the measurements not included in the DES Y3 cosmological analyses (grey regions in Fig. \ref{fig:sc}).}

\minorsection{Second and Third Map Moments}
Given current measurement uncertainties, the impact of source clustering on second moments is negligible (first row of Fig. \ref{fig:sc}), in line with the findings of \cite{Krause2021}. It only slightly dampens the signal at small scales and in moments that include a low redshift bin, for both `auto' and `cross' moments. 
For third moments the impact is more dramatic (second row of Fig. \ref{fig:sc}), particularly for moments that include low redshift bins. The data clearly follow better the simulations with source clustering, and the difference between the two sets of simulations is often significantly larger than measurement uncertainties. 

Most of the effect induced by source clustering is due to a non-zero correlation between the convergence field and the noise. The effect of source clustering for a mock sample with no shape noise is significantly smaller (but does not vanish completely, see Fig. \ref{fig:sc_2}). The non-zero noise-signal correlation follows from the noise modulation introduced in Eq. \ref{eq:sc_pixel}, and it is a consequence of the map-making procedure. This can also be tested in data by looking at third moments that combine the noisy convergence maps and `noise-only' maps created by randomly rotating the galaxy ellipticities of the shape catalogue. The rotation erases the shear signal but preserves the source clustering modulation of the noise. In simulations, we find that while moments of the form $\avg{{\kappa}^2{\kappa}_{{\rm N}}}$ or $\avg{{\kappa}_{{\rm N}}^3}$ are consistent with zero within uncertainties, $\avg{{\kappa}{\kappa}^2_{{\rm N}}}$ are not (in the presence of source clustering).  This is shown in the third row of Fig. \ref{fig:sc}, where simulations with source clustering provide a good match to the data. That $\avg{{\kappa}{\kappa}^2_{{\rm N}}}$ is non-zero was already noted in \cite{G20,moments2021}, although the nature of the effect was not then understood. To compare the measurements to theory predictions, the authors of those papers subtracted $\avg{{\kappa}{\kappa}^2_{{\rm N}}}$ from the estimated third moments $\langle{\kappa_{\rm obs}}^3\rangle$. The result of this procedure is shown in the fourth row of Fig. \ref{fig:sc}; the impact of source clustering is greatly minimised, {although the measurement errors are now larger}. This procedure completely removes the contribution due to the non-zero correlation between the convergence field and the noise, and leaves the part of the effect associated with the modification of the average shear signal in the pixels, which is sub-dominant. Using the simulations produced in this work, we verified that the scale cut adopted in the \cite{G20,moments2021} analysis, in combination with the subtraction of $\avg{{\kappa}{\kappa}^2_{{\rm N}}}$ terms, makes the analysis robust against source clustering effects (neglecting source clustering effects produces only a $0.08 \sigma$ shift in the marginalised two dimensional posterior of $\Omega_{\rm m}$ and $S_8$).

\minorsection{Peaks}
The fifth row of Fig. \ref{fig:sc} shows the impact of source clustering on the peak count function. We show the measurements only for the smoothing scale $\theta_0 = 13.2$ arcmin, intermediate among the several smoothing scales included in the DES Y3 peaks analysis in \cite{Zuercher2021b}; the trend with scales (not shown here) and redshift is similar to the moments case, i.e. the difference between the two sets of simulations increases with smaller smoothing scales and when low redshift bins are considered. Noise and signal are non-trivially mixed together due to the strong non-linearity of the peak function, and so, unlike the moments case, we did not try to create a procedure to minimise the impact of source clustering, nor did we try to single out the effects due to the extra noise-signal correlations. We found that for peaks statistic the effect is less striking than the moments case. We verified that for the scales considered in the analysis by \cite{Zuercher2021}, i.e. $[7.9, 31.6]$ arcmin, the difference between two simulated data vectors with and without source clustering is small enough to not bias the cosmological inference (neglecting source clustering effects produces only a $0.18 \sigma$ shift in the marginalised two dimensional posterior of $\Omega_{\rm m}$ and $S_8$).


\minorsection{WPH}
The last row of Fig. \ref{fig:sc} shows the WPH statistics obtained using one of the noisy convergence maps and one of the noise-only maps. We do not show the harmonics obtained using only the noisy convergence maps because a cosmological analysis using the measurements is currently underway (Gatti et al., in prep.); {since the measurements are blinded, we cannot compare them to simulations.} These statistics are consistent with zero in the absence of source clustering; however, {we detect a clear signal in data due to noise-signal correlations}, and this is well reproduced by the simulations with source clustering.

\minorsection{Significance}
Using the moments and the WPH coefficients, we can construct two null-tests for source clustering. {The $C01$ coefficients of the WPH statistics of noisy convergence maps and noise only maps are expected to be zero in the absence of source clustering (consistent with the simulation without source clustering). Using this null-test for the bins combination (3,1), we find a $p$-value for our observed $\chi^2$ of {$p=3\times10^{-11}$, which corresponds to 6.5 $\sigma$ significance.}  This result assumes a mean-zero Gaussian likelihood with covariance matrix $\Sigma$ estimated from simulations with no source clustering}, where $\chi^2 = d^\textrm{T} \Sigma^{-1} d$ with measured observable vector $d$. The same null-test for the third moment $\avg{{\kappa}{\kappa}^2_{{\rm N}}}$ for the bins combination (3,1,1) yields $p=4\times10^{-3}$ (2.6 $\sigma$). No trivial null-test can be constructed with the peaks statistics. 

Finally, we note that the magnitude of the source clustering effect also depends on the clustering properties of the source sample (e.g. the source galaxy-matter bias, Fig \ref{fig:sc_2}), which should be marginalised over when analysing map-based weak lensing higher-order statistics.
\section{Discussion and Conclusion}

We have demonstrated the impact of source galaxy clustering on map-based higher-order summary statistics of weak gravitational lensing observables. Source clustering affects the mean shear field estimated from galaxy catalogues, as the noise-free lensing signal is modulated by a different effective redshift distribution; moreover, it induces a strong correlation between a pixel's shear signal and its noise properties. The latter effect is the dominant one in map-based higher-order statistics. Using simulations with galaxies that either trace or do not trace the underlying density field, we show that the effect induced in the signals of common higher-order statistics can exceed the current measurement uncertainties, depending on the choice of scale cut and of summary statistic redshift range. We find that third moments and wavelet phase harmonic coefficients are the most affected ones, whereas peak counts are less affected. Source clustering effects are larger at small scales and for statistics applied to combinations of low and high redshift samples, and diminish at high redshift.

Further, we have shown a clear source clustering feature using Dark Energy Survey Year 3 data.  Due to the induced correlation between the shear signal and the noise properties of the maps, third moments combining the noisy convergence maps and `noise-only' maps no longer vanish. We detected a similar feature at high statistical significance for wavelet phase harmonics. Mocks with source clustering were well able to reproduce these features; mocks without source clustering provided a poor fit to the data ($p$-values of 4e-3 for third moments and 3e-11 for wavelet phase harmonics).

{Cosmological analyses using map-based higher-order statistics have two strategies for dealing with source clustering: either minimise its effect by introducing \textit{ad-hoc} scale cuts and/or de-noising procedures, or fully forward model it, incorporating it into simulations. This work presents a recipe for efficiently incorporating source clustering effects into simulations, and also shows how to minimise the impact of source clustering for third moments using a de-noising procedure. If left unaccounted for, or if not tested, this effect could impact cosmological inference made with statistics using weak gravitational lensing observables, especially map-based higher-order statistics (including ones not considered here, e.g. scattering transforms, deep learning summary statistics, Minkowski functionals, etc.). In the case of the DES Y3 higher-order statistics analyses -- moments \citep{moments2021} and peaks \citep{Zuercher2022} -- we verified that the scale cuts and de-noising procedures adopted were sufficient to render this effect negligible. }

{Other effects could cause noise-signal correlations in map-based estimators, e.g. any selection effect depending on the local value of the matter and shear fields modulating the source number density. Source magnification induces an extra modulation proportional to $1 + \kappa(p,s)$, however our tests shows this to be negligible (owing to a lower signal amplitude compared to the density field). Blending effects are also likely negligible, as they are expected to affect only a small fraction of the sample. In general, any deviation from the simple $1+b_g \delta(p,s)$ modulation considered here would lead to a specific redshift evolution and/or amplitude signature in the measurements, and we do not see this. Other astrophysical effects such as intrinsic alignment and baryonic feedback can impact $\gamma(s,p)$ and $\delta(s,p)$, but they do not directly modulate the number of galaxies. They could, however, enhance the source clustering effects: intrinsic alignment, in particular, is a local effect modulated by the same density fluctuations that modulate the source clustering \citep{Blazek2019}, and hence it could boost the amplitude of the noise-signal correlations.}

{This work focused on map-based statistics. Source clustering is expected to affect catalogue-based statistics differently: there should be no noise-signal contributions (as these are due to averaging the shear in pixels before estimating the summary statistics), but sources would still be preferentially sampled in regions with high shear/convergence. The impact is thus expected to be smaller; we leave this investigation to future works.}




\bibliography{bibliography,des_y3kp}
\bibliographystyle{mn2e_2author_arxiv_amp.bst}


\newpage
\appendix
\input{Appendix}

\input{aff.tex}


\label{lastpage}
\end{document}

%% file: authors_MNRAS.tex
\author[M. Gatti et al.]{
\parbox{\textwidth}{
\large{M.~Gatti$^{1\star}$,  
N.~Jeffrey$^{2}$  
L.~Whiteway$^{2}$
V.~Ajani$^{3}$,
T.~Kacprzak$^{3}$,
D.~Zürcher$^{3}$,
C.~Chang$^{4,5}$, 
B.~Jain$^{1}$, 
J.~Blazek$^{6}$,
E.~Krause$^{7}$
% ...
\begin{center} (DES Collaboration) \end{center}
}
}}

%% file: Appendix.tex
\section{SUPPLEMENTARY MATERIAL: IMPLEMENTATION OF THE SUMMARY STATISTICS }\label{sect:summary_stats}

We provide here additional details concerning the implementation of the summary statistics used in this work. 

\minorsection{Second and Third Map Moments}
The implementation of second and third map moments follows \cite{moments2021}. We first smooth the maps using a top-hat filter with different smoothing scales. 
%
We consider ten equally (logarithmic) spaced smoothing scales $\theta_0$ between 3.2 and 200 arcmin.  The second and third moments of the smoothed maps are computed as follows:
\begin{equation}
\label{eq:second_moments}
\avg{\kappa^2_{\theta_0}}^{i,j} = \frac{1}{N_{\mathrm{tot}}}\sum_{\mathrm{pix}}^{N_{\rm tot}} \kappa_{\theta_0,\mathrm{pix}}^i \kappa_{\theta_0,\mathrm{pix}}^j,
\end{equation}
\begin{equation}
\avg{\kappa^3_{\theta_0}}^{i,j,k} = \frac{1}{N_{\mathrm{tot}}}\sum_{\mathrm{pix}}^{N_{\rm tot}} \kappa_{\theta_0,\mathrm{pix}}^i \kappa_{\theta_0,\mathrm{pix}}^j \kappa_{\theta_0, \mathrm{pix}}^k.
\end{equation}
We can only estimate \textit{noisy} realisations of the weak lensing mass maps: $\kappa_{{\rm obs}} = \kappa + \kappa_{{\rm N}}$. Any statistic measured with data will include noise contributions  \citep{VanWaerbeke2013}, for example:
\begin{multline}
\label{eq:deno3}
\langle{\kappa_{\rm obs}}^3\rangle^{i,j,k} =  \langle \kappa^3\rangle^{i,j,k} +  \langle \kappa^3_{{\rm N}}\rangle^{i,j,k} + \\ \left[\langle  \kappa \kappa_{{\rm N}} \kappa_{{\rm N}} \rangle^{i,j,k} + \langle  \kappa \kappa \kappa_{{\rm N}}  \rangle^{i,j,k} + {\rm cycl.} \right],
\end{multline}
where `$\rm{cycl.}$' refers to the cyclic permutation of the tomographic bin indexes $i$, $j$, $k$ for the terms in parenthesis. Many of these terms are expected to be affected by source clustering. Most strikingly, certain combinations of terms that would otherwise be expected to be zero can become non-zero due to source clustering (e.g. $\avg{\kappa \kappa_{\rm N}^2}$ or $\avg{\kappa_{\rm obs} \kappa_{\rm N}^2} = \avg{\kappa \kappa_{\rm N}^2} + \avg{\kappa_{\rm N}^3}$). Note we can estimate $\kappa_{\rm N}$ from noise-only shear maps.



\minorsection{Peaks}
We follow the implementation of peak counts in \cite{Zuercher2021b}. We smooth the maps with Gaussian filters of different scales (12 scales with full-width-half-maximum between 2.6 and 31.6 arcmin), and we use $\sim$ 15 equally spaced thresholds in the value of $\kappa$. 

Peaks are detected in the maps corresponding to the four tomographic bins (`auto' peaks). Peaks are also detected in  maps obtained by combining two convergence maps from different tomographic bins  (`cross' peaks), following the procedure outlined in \cite{Zuercher2021}. As standard practice, peak counts obtained from noise-only maps are subtracted off the measurements. This subtraction is not guaranteed to completely remove the noise-only contribution from the measurement, due the non-linearity of the peak function. 



\minorsection{WPH}
Our implementation of the WPH follows \cite{Allys2020}. 
We use the package \href{https://github.com/bregaldo/pywph}{pyWPH}~\citep{bruno_denoising} to measure the WPH statistics from flat-sky projections of the weak lensing mass maps. We first cut multiple square patches out of the DES footprint; the patches are made of 64$\times$64 pixels, with a pixel scale of 6.8 arcmin. Each patch is smoothed by a \textit{bump steerable wavelet} ${\psi}_{j,\ell}$, where $j$ specifies the spatial frequency of the order of $2^{j+1}$ pixels, and $\ell$ a rotation angle $2\pi \ell / L$ (see \citealt{Allys2020} for detailed definitions). We consider $j = {0,1,2,3}$ (with $j=3$ corresponding to a frequency of $\approx109$ arcmin), and  $\ell = {0,1,2}$ with $L = 3$.

%
%

We apply a non-linear operation to the smoothed field that allows the capturing of interactions between scales and that provides access to non-Gaussian features of the field using second moments.\footnote{While the second moments of a wavelet transformed field depend only on the power spectrum, the second moments of a  wavelet transformed field that has undergone a non-linear transformation depend on the field's higher order moments.} Define the \textit{phase harmonic of order $p$} as \citep{Mallat2020}:
\begin{equation}
    {\rm PH} (r e^{\textrm{i} \theta},p) \equiv r e^{\textrm{i} p \theta}.
\end{equation}
Many summary statistics can be constructed computing the second moments of two transformed and smoothed fields \citep{Allys2020}. In this work, since we are not interested in constraining cosmological parameters but rather in showcasing the impact of SC, we limit ourselves to one summary statistic: \begin{multline}
    C01^{i,k}_{j_1,j_2\neq j_1} \equiv \frac{1}{N_{\mathrm{tot}}} \sum_{\mathrm{pix}}^{N_{\rm tot}} \sum_{\ell}^{L} {\rm PH}(\kappa_{{j_1,\ell}}^i,0) {\rm PH} (\kappa_{{j_2,\ell}}^k,1) = \\ \frac{1}{N_{\mathrm{tot}}} \sum_{\mathrm{pix}}^{N_{\rm tot}} \sum_{\ell}^{L}|\kappa_{{j_1,\ell}}^i| \kappa_{{j_2,\ell}}^k,
\end{multline}
\noindent where $\kappa_{{j,\ell}}^i$ is the map of the $i$-th tomographic bin smoothed by the filter ${\psi}_{j,\ell}$ and where we have considered the case $j_1 \neq j_2$.

\section{SUPPLEMENTARY MATERIAL: SOURCE CLUSTERING, SHAPE NOISE and GALAXY-MATTER BIAS}\label{sect:summary_stats_2}

\begin{figure}
\begin{center}
\includegraphics[width=0.48 \textwidth]{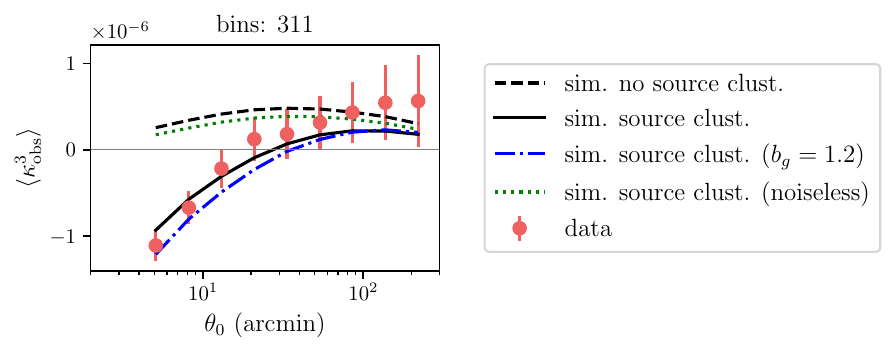}
\end{center}
\caption{Third moments as measured in simulations using different mocks: with no source clustering, with source clustering, with source clustering  but assuming a stronger linear clustering for the source galaxies (i.e., assuming a galaxy-matter bias of $b_g=1.2$).}
\label{fig:sc_2}
\end{figure}

We show in Fig. \ref{fig:sc_2} the effect of source clustering on third moments for a mock sample with no shape noise. The effect is significantly smaller then in the case with shape noise (but does not vanish completely). In the same Figure we also show that the magnitude of the source clustering effect depends on the clustering properties of the source sample (e.g. the sources galaxy-matter bias), 
which should be marginalised over when analysing map-based weak lensing higher-order statistics.

%% file: aff.tex
\section{Affiliations}\label{sec:affiliations}
{$^{\star}$ E-mail: marcogatti29@gmail.com}\\
$^{1}$ Department of Physics and Astronomy, University of Pennsylvania, Philadelphia, PA 19104, USA\\
$^{2}$ Department of Physics \& Astronomy, University College London, Gower Street, London, WC1E 6BT, UK\\
$^{3}$ Department of Physics, ETH Zurich, Wolfgang-Pauli-Strasse 16, CH-8093 Zurich, Switzerland \\
$^{4}$ Department of Astronomy and Astrophysics, University of Chicago, Chicago, IL 60637, USA \\
$^{5}$ Kavli Institute for Cosmological Physics, University of Chicago, Chicago, IL 60637, USA \\
$^{6}$ Department of Physics, Northeastern University, Boston, MA 02115, USA \\
$^{7}$  Department of Astronomy/Steward Observatory, University of Arizona, 933 North Cherry Avenue, Tucson, AZ 85721-0065, USA \\
$^{8}$  Argonne National Laboratory, 9700 South Cass Avenue, Lemont, IL 60439, USA\\
$^{9}$  Institute of Astronomy, University of Cambridge, Madingley Road, Cambridge CB3 0HA, UK\\
$^{10}$ Kavli Institute for Cosmology, University of Cambridge, Madingley Road, Cambridge CB3 0HA, UK\\
$^{11}$ Physics Department, 2320 Chamberlin Hall, University of Wisconsin-Madison, 1150 University Avenue Madison, WI  53706-1390\\
$^{12}$ Department of Physics, Carnegie Mellon University, Pittsburgh, Pennsylvania 15312, USA \\
$^{13}$ Department of Physics, Duke University Durham, NC 27708, USA\\
$^{14}$  NASA Goddard Space Flight Center, 8800 Greenbelt Rd, Greenbelt, MD 20771, USA\\
$^{15}$ Kavli Institute for Particle Astrophysics \& Cosmology, P. O. Box 2450, Stanford University, Stanford, CA 94305, USA\\
$^{16}$ Lawrence Berkeley National Laboratory, 1 Cyclotron Road, Berkeley, CA 94720, USA\\
$^{17}$ Fermi National Accelerator Laboratory, P. O. Box 500, Batavia, IL 60510, USA \\
$^{18}$ NSF AI Planning Institute for Physics of the Future, Carnegie Mellon University, Pittsburgh, PA 15213, USA\\
$^{19}$ Universit\'e Grenoble Alpes, CNRS, LPSC-IN2P3, 38000 Grenoble, France\\
$^{20}$ Department of Physics and Astronomy, University of Waterloo, 200 University Ave W, Waterloo, ON N2L 3G1, Canada\\
$^{21}$ Jet Propulsion Laboratory, California Institute of Technology, 4800 Oak Grove Dr., Pasadena, CA 91109, USA\\
$^{22}$ SLAC National Accelerator Laboratory, Menlo Park, CA 94025, USA\\
$^{23}$ University Observatory, Faculty of Physics, Ludwig-Maximilians-Universit\"at, Scheinerstr. 1, 81679 Munich, Germany\\
$^{24}$ Center for Astrophysical Surveys, National Center for Supercomputing Applications, 1205 West Clark St., Urbana, IL 61801, USA\\
$^{25}$ Department of Astronomy, University of Illinois at Urbana-Champaign, 1002 W. Green Street, Urbana, IL 61801, USA\\
$^{26}$ School of Physics and Astronomy, Cardiff University, CF24 3AA, UK\\
$^{27}$ Department of Astronomy, University of Geneva, ch. d'\'Ecogia 16, CH-1290 Versoix, Switzerland \\
$^{28}$ Department of Applied Mathematics and Theoretical Physics, University of Cambridge, Cambridge CB3 0WA, UK \\
$^{29}$ Department of Physics, Stanford University, 382 Via Pueblo Mall, Stanford, CA 94305, USA \\
$^{30}$ Instituto de F\'isica Gleb Wataghin, Universidade Estadual de Campinas, 13083-859, Campinas, SP, Brazil\\
$^{31}$ Department of Physics, University of Genova and INFN, Via Dodecaneso 33, 16146, Genova, Italy\\
$^{32}$ Jodrell Bank Center for Astrophysics, School of Physics and Astronomy, University of Manchester, Oxford Road, Manchester, M13 9PL, UK\\
$^{33}$ Centro de Investigaciones Energ\'eticas, Medioambientales y Tecnol\'ogicas (CIEMAT), Madrid, Spain\\
$^{34}$ Brookhaven National Laboratory, Bldg 510, Upton, NY 11973, USA\\
$^{35}$  Department of Physics and Astronomy, Stony Brook University, Stony Brook, NY 11794, USA\\
$^{36}$ Department of Physics, Duke University Durham, NC 27708, USA\\
$^{37}$  Institut de Recherche en Astrophysique et Plan\'etologie (IRAP), Universit\'e de Toulouse, CNRS, UPS, CNES, 14 Av. Edouard Belin, 31400 Toulouse, France\\
$^{38}$ Institut d'Estudis Espacials de Catalunya (IEEC), 08034 Barcelona, Spain\\
$^{39}$ Institute of Space Sciences (ICE, CSIC),  Campus UAB, Carrer de Can Magrans, s/n,  08193 Barcelona, Spain\\
$^{40}$ Excellence Cluster Origins, Boltzmannstr.\ 2, 85748 Garching, Germany\\
$^{41}$ Max Planck Institute for Extraterrestrial Physics, Giessenbachstrasse, 85748 Garching, Germany\\
$^{42}$ Universit\"ats-Sternwarte, Fakult\"at f\"ur Physik, Ludwig-Maximilians Universit\"at M\"unchen, Scheinerstr. 1, 81679 M\"unchen, Germany\\
$^{43}$ Cerro Tololo Inter-American Observatory, NSF's National Optical-Infrared Astronomy Research Laboratory, Casilla 603, La Serena, Chile\\
$^{44}$ Department of Astronomy, University of Michigan, Ann Arbor, MI 48109, USA\\
$^{45}$ Institute for Astronomy, University of Edinburgh, Edinburgh EH9 3HJ, UK\\
$^{46}$ Instituto de Astrofisica de Canarias, E-38205 La Laguna, Tenerife, Spain\\
$^{47}$ Laborat\'orio Interinstitucional de e-Astronomia - LIneA, Rua Gal. Jos\'e Cristino 77, Rio de Janeiro, RJ - 20921-400, Brazil\\
$^{48}$ Universidad de La Laguna, Dpto. Astrofísica, E-38206 La Laguna, Tenerife, Spain\\
$^{49}$ Institute of Cosmology and Gravitation, University of Portsmouth, Portsmouth, PO1 3FX, UK\\
$^{50}$ Center for Astrophysics $\vert$ Harvard \& Smithsonian, 60 Garden Street, Cambridge, MA 02138, USA\\
$^{51}$ Santa Cruz Institute for Particle Physics, Santa Cruz, CA 95064, USA\\
$^{52}$ CNRS, UMR 7095, Institut d'Astrophysique de Paris, F-75014, Paris, France\\
$^{53}$ Sorbonne Universit\'es, UPMC Univ Paris 06, UMR 7095, Institut d'Astrophysique de Paris, F-75014, Paris, France\\
$^{54}$ Computer Science and Mathematics Division, Oak Ridge National Laboratory, Oak Ridge, TN 37831\\
$^{55}$ Department of Physics, University of Michigan, Ann Arbor, MI 48109, USA\\
$^{56}$ Institute of Theoretical Astrophysics, University of Oslo. P.O. Box 1029 Blindern, NO-0315 Oslo, Norway\\
$^{57}$ George P. and Cynthia Woods Mitchell Institute for Fundamental Physics and Astronomy, and Department of Physics and Astronomy, Texas A\&M University, College Station, TX 77843,  USA\\
$^{58}$ Institut de F\'{\i}sica d'Altes Energies (IFAE), The Barcelona Institute of Science and Technology, Campus UAB, 08193 Bellaterra (Barcelona) Spain\\
$^{59}$ Instituto de Fisica Teorica UAM/CSIC, Universidad Autonoma de Madrid, 28049 Madrid, Spain\\
$^{60}$ Center for Cosmology and Astro-Particle Physics, The Ohio State University, Columbus, OH 43210, USA\\
$^{61}$ Department of Physics, The Ohio State University, Columbus, OH 43210, USA\\
$^{62}$ Australian Astronomical Optics, Macquarie University, North Ryde, NSW 2113, Australia\\
$^{63}$ Lowell Observatory, 1400 Mars Hill Rd, Flagstaff, AZ 86001, USA\\
$^{64}$ Hamburger Sternwarte, Universit\"{a}t Hamburg, Gojenbergsweg 112, 21029 Hamburg, Germany\\
$^{65}$ School of Physics and Astronomy, University of Southampton,  Southampton, SO17 1BJ, UK\\
$^{66}$ Instituci\'o Catalana de Recerca i Estudis Avan\c{c}ats, E-08010 Barcelona, Spain\\
$^{67}$ Observat\'orio Nacional, Rua Gal. Jos\'e Cristino 77, Rio de Janeiro, RJ - 20921-400, Brazil\\
$^{68}$ Physics Department, William Jewell College, Liberty, MO, 64068\\
$^{69}$ School of Mathematics and Physics, University of Queensland,  Brisbane, QLD 4072, Australia\\
$^{70}$ Department of Physics, IIT Hyderabad, Kandi, Telangana 502285, India\\
$^{71}$ School of Mathematics and Physics, University of Queensland,  Brisbane, QLD 4072, Australia\\

%% file: main.bbl
\begin{thebibliography}{42}
\providecommand{\natexlab}[1]{#1}
\providecommand{\url}[1]{\texttt{#1}}
\providecommand{\urlprefix}{URL }
\providecommand{\eprint}[1][]{\url{#1}}

\bibitem[{{Allys} et~al.(2020)}]{Allys2020}
{Allys}, E., et~al., 2020, \prd, 102, 10, 103506

\bibitem[{{Bartelmann} \& {Schneider}(2001)}]{Bartelmann2001}
{Bartelmann}, M., {Schneider}, P., 2001, \physrep, 340, 4-5, 291

\bibitem[{{Blazek} et~al.(2019)}]{Blazek2019}
{Blazek}, J., et~al., 2019, \prd, 100, 10, 103506

\bibitem[{{Cardone} et~al.(2013){Cardone} \& {Camera} et~al.}]{Cardone2013}
{Cardone}, V.~F., {Camera}, S., {Mainini}, R., et~al., 2013, \mnras, 430, 4,
  2896

\bibitem[{{Chang} et~al.(2018){Chang} \& {Pujol} et~al.}]{Chang2018}
{Chang}, C., {Pujol}, A., {Mawdsley}, B., et~al., 2018, \mnras, 475, 3165

\bibitem[{{Cheng} et~al.(2020){Cheng} \& {Ting} \& {M{\'e}nard} \&
  {Bruna}}]{Cheng2020}
{Cheng}, S., {Ting}, Y.-S., {M{\'e}nard}, B., {Bruna}, J., 2020, \mnras, 499,
  4, 5902

\bibitem[{{Fluri} et~al.(2019){Fluri} \& {Kacprzak} et~al.}]{Fluri2019}
{Fluri}, J., {Kacprzak}, T., {Lucchi}, A., et~al., 2019, \prd, 100, 6, 063514

\bibitem[{{Fosalba} et~al.(2015)}]{Fosalba2015}
{Fosalba}, P., et~al., 2015, \mnras, 447, 2, 1319

\bibitem[{{Gatti} et~al.(2020){Gatti} \& {Chang} et~al.}]{G20}
{Gatti}, M., {Chang}, C., {Friedrich}, O., et~al., 2020, \mnras, 498, 3, 4060

\bibitem[{{Gatti} et~al.(2022){Gatti} \& {Jain} et~al.}]{moments2021}
{Gatti}, M., {Jain}, B., {Chang}, C., et~al., 2022, \prd, 106, 8, 083509

\bibitem[{{Gatti} et~al.(2021){Gatti} \& {Sheldon} et~al.}]{y3-shapecatalog}
{Gatti}, M., {Sheldon}, E., {Amon}, A., et~al., 2021, \mnras, 504, 3, 4312

\bibitem[{{G{\'o}rski} et~al.(2005){G{\'o}rski} \& {Hivon} et~al.}]{GORSKI2005}
{G{\'o}rski}, K.~M., {Hivon}, E., {Banday}, A.~J., et~al., 2005, \apj, 622, 759

\bibitem[{{Huff} \& {Mandelbaum}(2017)}]{HuffMcal2017}
{Huff}, E., {Mandelbaum}, R., 2017, arXiv e-prints, 1702.02600

\bibitem[{{Jeffrey} et~al.(2021{\natexlab{a}}){Jeffrey} \& {Alsing} \&
  {Lanusse}}]{jeffrey_lfi}
{Jeffrey}, N., {Alsing}, J., {Lanusse}, F., 2021{\natexlab{a}}, \mnras, 501, 1,
  954

\bibitem[{{Jeffrey} et~al.(2021{\natexlab{b}}){Jeffrey} \& {Gatti}
  et~al.}]{y3-massmapping}
{Jeffrey}, N., {Gatti}, M., {Chang}, C., et~al., 2021{\natexlab{b}}, \mnras,
  505, 3, 4626

\bibitem[{{Jeffrey} et~al.(2022)}]{jeffrey_lfi_wph}
{Jeffrey}, N., et~al., 2022, \mnras, 510, 1, L1

\bibitem[{{Kacprzak} et~al.(2016){Kacprzak} \& {Kirk} et~al.}]{Kacprzak2016}
{Kacprzak}, T., {Kirk}, D., {Friedrich}, O., et~al., 2016, \mnras, 463, 3653

\bibitem[{{Kaiser} \& {Squires}(1993)}]{KaiserSquires}
{Kaiser}, N., {Squires}, G., 1993, \apj, 404, 441

\bibitem[{{Krause} et~al.(2021)}]{Krause2021}
{Krause}, E., et~al., 2021, arXiv e-prints, 2105.13548

\bibitem[{{Liu} et~al.(2015)}]{Liu2015}
{Liu}, J., et~al., 2015, \prd, 91, 6, 063507

\bibitem[{{Lu} et~al.(2023){Lu} \& {Haiman} \& {Li}}]{Lu2023}
{Lu}, T., {Haiman}, Z., {Li}, X., 2023, \mnras, 521, 2, 2050

\bibitem[{{MacCrann} et~al.(2022)}]{y3-imagesims}
{MacCrann}, N., et~al., 2022, \mnras, 509, 3, 3371

\bibitem[{Mallat(2016)}]{Mallat_2016}
Mallat, S., 2016, Philosophical Transactions of the Royal Society A:
  Mathematical, Physical and Engineering Sciences, 374, 2065, 20150203

\bibitem[{{Mallat} et~al.(2020)}]{Mallat2020}
{Mallat}, S., et~al., 2020, Information and Inference: A Journal of the IMA, 9,
  3, 721, ISSN 2049-8772

\bibitem[{{Martinet} et~al.(2018)}]{Martinet2018}
{Martinet}, N., et~al., 2018, \mnras, 474, 1, 712

\bibitem[{{Myles} et~al.(2021){Myles} \& {Alarcon} et~al.}]{y3-sompz}
{Myles}, J., {Alarcon}, A., {Amon}, A., et~al., 2021, \mnras, 505, 3, 4249

\bibitem[{{Peel} et~al.(2018)}]{Peel2018}
{Peel}, A., et~al., 2018, \aap, 619, A38

\bibitem[{{Petri} et~al.(2015)}]{Petri2015}
{Petri}, A., et~al., 2015, \prd, 91, 10, 103511

\bibitem[{Potter et~al.(2017)}]{potter2017pkdgrav3}
Potter, D., et~al., 2017, Computational Astrophysics and Cosmology, 4, 1, 2

\bibitem[{{Pyne} \& {Joachimi}(2021)}]{Pyne2021}
{Pyne}, S., {Joachimi}, B., 2021, \mnras, 503, 2, 2300

\bibitem[{{Regaldo-Saint Blancard} et~al.(2021)}]{bruno_denoising}
{Regaldo-Saint Blancard}, B., et~al., 2021, \aap, 649, L18

\bibitem[{{Schmidt} et~al.(2009)}]{Schmidt2009}
{Schmidt}, F., et~al., 2009, \apj, 702, 1, 593

\bibitem[{{Schneider} et~al.(2002){Schneider} \& {van Waerbeke} \&
  {Mellier}}]{Schneider2002}
{Schneider}, P., {van Waerbeke}, L., {Mellier}, Y., 2002, \aap, 389, 729

\bibitem[{{Sheldon} \& {Huff}(2017)}]{SheldonMcal2017}
{Sheldon}, E.~S., {Huff}, E.~M., 2017, \apj, 841, 24

\bibitem[{{Valageas}(2014)}]{Valageas2014}
{Valageas}, P., 2014, \aap, 561, A53

\bibitem[{{Van Waerbeke} et~al.(2013){Van Waerbeke} \& {Benjamin}
  et~al.}]{VanWaerbeke2013}
{Van Waerbeke}, L., {Benjamin}, J., {Erben}, T., et~al., 2013, \mnras, 433,
  3373

\bibitem[{{Vicinanza} et~al.(2016)}]{Vicinanza2016}
{Vicinanza}, M., et~al., 2016, arXiv e-prints, arXiv:1606.03892

\bibitem[{{Vicinanza} et~al.(2018)}]{Vicinanza2018}
{Vicinanza}, M., et~al., 2018, \prd, 97, 2, 023519

\bibitem[{Zhang \& Mallat(2021)}]{Zhang2019}
Zhang, S., Mallat, S., 2021, ACHA, 53, 199, ISSN 1063-5203

\bibitem[{{Z{\"u}rcher} et~al.(2021)}]{Zuercher2021}
{Z{\"u}rcher}, D., et~al., 2021, \jcap, 2021, 1, 028

\bibitem[{{Z{\"u}rcher} et~al.(2022{\natexlab{a}})}]{Zuercher2021b}
{Z{\"u}rcher}, D., et~al., 2022{\natexlab{a}}, \mnras, 511, 2, 2075

\bibitem[{{Z{\"u}rcher} et~al.(2022{\natexlab{b}})}]{Zuercher2022}
{Z{\"u}rcher}, D., et~al., 2022{\natexlab{b}}, arXiv e-prints, arXiv:2206.01450

\end{thebibliography}
